# Search for the potential electromagnetic counterparts of neutrino events in SDSS galaxies at z<0.1

[1]Sergijenko O., PhD in Phys. & Math., Senior Researcher
https://orcid.org/0000-0002-9212-7118
[1]Vavilova I.B., Dr. Sci. Hab. in Phys. & Math., Prof., Head of Department
ORCID https://doi.org/0000-0002-5343-1408
[1,2] Izviekova I.O., Junior Researcher, Post-graduate student
ORCID  https://orcid.org/0009-0009-4307-0627
[1,3] Karakuts D.R., Junior Researcher, Post-graduate student
ORCID https://orcid.org/0009-0001-8219-0527

[1] Main Astronomical Observatory of the NAS of Ukraine
27 Akademik Zabolotnyi St., Kyiv 03143 Ukraine
[2] ICAMER Observatory of the NAS of Ukraine
27 Akademik Zabolotnyi St., Kyiv 03143 Ukraine
[3] Institute of Physics of the NAS of Ukraine,
46 e Nauka Av., Kyiv 03028 Ukraine

**Abstract**. Identification of the electromagnetic emission in coincidence with the high-energy neutrinos is fundamentally important in the multimessenger astronomy. Such observations are essential for constraining the source localization, determining the source type and understanding the emission mechanisms. Generally, they require following up a neutrino alert (there are 2 streams of alerts for IceCube: Gold, having at least 50 % probability of astrophysical origin, and Bronze, with such probaility of at least 30 %) with an electromagnetic facility (with the primary interest in X and gamma rays), but also involve an electromagnetic monitoring of the hot-spots (points exceeding the instrument sensitivity) in the skymap provided by IceCube. Alternative approach consists in performing the correlation analysis across the available neutrino events and catalogs of sources.

We searched for spatial coincidence between the galaxies from SDSS and the high-energy neutrino events. The IceCube Gold alerts and the neutrino-electromagnetic coincidence events from AMON (Astrophysical Multimessenger Observatory Network) identified until the end of September 2025 are considered. Galaxies from the Morphological catalog of galaxies at 0.02<z<0.1 (including 315,776 objects from SDSS DR9 with the absolute stellar magnitudes in the range of $-24^m < M^r < -13^m$) are examined. Among 59 IceCube Gold alerts we ound 3 with only 1 galaxy (SDSS J231231.52+033415.1) within the 50 % containment radius. Among 24 neutrino-electromagnetic coincidence events there are also 3 with only 1 galaxy (SDSS J220711.14+122535.9) within the 50 % containment radius. These 6 galaxies are the most promising candidates for host galaxies of the neutrino sources. We summarize their available multiwavelength data and light cuves raken from ZTF for period 2018–2025.

# 1. Introduction

Astrophysical sources capable of hadronic acceleration to relativistic energies are considered to be the potential sources of astrophysical neutrinos which, unlike photons, are able to travel through dense astrophysical environments and over cosmological distances without the Extragalactic Background Light (EBL) absorption and provide unambiguous tracers of hadronic acceleration. Since 2013, when the IceCube Neutrino Observatory eported the presence of an all-sky, isotropic, high-energy neutrino flux of extraterrestrial origin [9], the sources responsible for the measured astrophysical neutrino flux and their identification are still under discussion. Currently only the following astronomical objects have been identified (all by the IceCube Collaboration) as high-energy neutrino emitters with a significance larger than $3\sigma$: the blazar TXS 0506+056 [2] at z = 0.3365, the local (z = 0.004) Seyfert 2 galaxy NGC 1068 [3] and the Galactic Plane [10]. There are also claims for neutrino excess from other Seyfert galaxies as well as from such classes of sources as an extended Cygnus Cocoon [13] and Tidal Disruption Events (TDEs)/TDE candidates identified by the Zwicky Transient Facility (ZTF) [16, 19].

In this context, we bote the Alert Management, Photometry, and Evaluation of Light curves (AMPEL) [14] as a multi-tier framework to analyze and act automatically on transient events. It could use IceCube alerts as native inputs and combine them statistically with optical streams (Zwicky Telescope Facility (ZTF), the Rubin Observatory LSST in the near future). In addition to real-time streaming, AMPEL can operate post-factum: archival IceCube notices/skymaps and curated event catalogs can be bulk-ingested, enabling retrospective cross-correlation and sensitivity studies.

At T0 tier (ingestion), a dedicated ingester normalizes each neutrino alert into AMPEL's schema: detection time, RA/Dec, skymap region, event class, energy proxy, and other fields.

At T1 tier (combine), optical alerts are grouped by astrophysical object. This tier prepares join keys (sky position, time window) and guarantees a consistent state even if the same data arrives multiple times. At T2 tier (analysis/classification), AMPEL tests each optical candidate against recent IceCube alerts with an unbinned spatio-temporal likelihood: evaluate the alert's directional PDF at the candidate position (or integrate over a small aperture), combine with a class-appropriate prior on $|\Delta t|$, and include event-quality terms (e.g., signalness or an energy proxy, when available). The resulting test statistic is calibrated by scrambling to obtain a false-alarm probability, yielding a ranked list of associations with uncertainties and full provenance. At T3 (reaction), policy thresholds trigger auditable actions: TNS updates, and community circulars, optionally conditioned on event class and localization precision. Because AMPEL preserves streams and configurations, we can reprocess back archives or adopt new priors. Alert database searches are done by AMPEL [8, 14], which allows combining sky localization of gravitational-wave candidates (LIGO/Virgo) and the coincident track-like neutrino event with ZTF reported by IceCube (see, also, review [6]). New three different AMPEL channels tuned for alert streams: to rapidly identify and process transients were recently proposed for public use [15].

Generally, searches for the electromagnetic (EM) counerparts of neutrino events and forecasts for the future EM facilities (e.g. for CTAO [17, 18]) aim to find a particular source. Our goal is to find potential host galaxies of such sources.

We describe the data in Section 2, results and lisrs of potential host galaxies are presemted in Section 3; discussion and conclusion are given in Section 4.

**2. Data mining**

Generally, searches for the electromagnetic (EM) counerparts of neutrino events aim to find a particular source. Our goal is to find potential host galaxies of such sources.

Within the IceCube real-time alert program of of neutrino events [1] (initiated in April 2016) that are likely to be of astrophysical origin, are reported via the General Coordinates Network (GCN; https://gcn.nasa.gov). On June 17, 2019 the updated alert system was introduced [5], in which all alerts are classified based on the likelihood of being from astrophysical origin. For the Gold alerts this probability is at least 50 %, while for Bronze alerts it is 30 %. Here we restrict consideration only to the Gold alerts[1]: 59 of them have been sent until the end of September 2025.

In order to perform a real-time correlation analysis of the high-energy EM and neutrino/gravitational wave signals the Astrophysical Multimessenger Observatory Network[2] (AMON) has been established [4]). We consider the events that are coincidences between neutrino detectors (mainly IceCube, but also ANTARES) and X or gamma ray detecting instruments. The 24 such events have been identified[3] until the end of September 2025.

We search for the spatial coincidence between the Gold neutrino alerts from IceCube and the galaxies from the image-based Morphological catalog of galaxies at $0.02<z<0.1$ [7] (available at [24]). We also search for the spatial coincidence between these galaxies and the coincidences between neutrino detectors and EM-detecting instruments. Neutrino events data lack the redshift information, therefore we can only analyze angular separations between events involving neutrino and galaxies.

The Morphological catalog of galaxies at $0.02<z<0.1$ contains of 315,776 galaxies with the absolute stellar magnitudes in the range of $-24^m < M_r < -13^m$ from the SDSS DR9 [24]. It was obtained by human labeling, multi-photometry supervised machine learning methods [22] and CNN classifier [11]. For the photometric binary morphological classification, the absolute magnitudes $M_u$, $M_g$, $M_r$, $M_i$, $M_z$; color indices $M_u-M_r$, $M_g-M_i$, $M_u-M_g$, $M_r-M_z$; and the inverse concentration index to the center R50/R90 were used. The supervised methods provide the accuracy of 96.4% for Support Vector Machine (96.1% early and 96.9% late types) and 95.5% for Random Forest (96.7% early and 92.8% late types). To obtain the CNN image-based classification of morphological classes and features, the dataset of Galaxy Zoo 2 [27] was applied as the training sample. The accuracy of CNN-classifier on the image-based morphological classes is as follows: cigar-shaped (75 %), completely round (83 %), round in-between (93 %), edge-on (93 %), spiral (96 %). As for the classification of galaxies by their detailed 32 structural morphological features, CNN model had the accuracy in the range of 83.3–99.4 % depending on features (bar, rings, number of spiral arms, mergers, dust lane, edge-on, etc.), a number of galaxies with the given feature in the inference dataset, and the galaxy image quality [23, 25].

---

[1] https://gcn.gsfc.nasa.gov/amon_icecube_gold_bronze_events.html
[2] https://gcn.gsfc.nasa.gov/amon.html
[3] https://gcn.gsfc.nasa.gov/amon_nu_em_coinc_events.html

## 3. Results

The obtained results of search between the IceCube Gold alerts and the number of potential host galaxies of electromagnetic counterparts within 50 % and 90 % containment radii are presented in Table 1. The 50% (90%) containment radius is a radius within which the probability of localization of a source is 0.5 (0.9), in arcmin. We did not consider corrections for the redshift if galaxies.

We have found that 30 (out of 59) neutrino events have no galaxies within their 90 % containment error radius (*italic* in Table 1). The largest number of galaxies within 90 % containment error radius of a neutrino event is 27487. Within 50 % containment error radius 33 neutrino events have no galaxies. The largest number of galaxies within 50 % containment error radius of a neutrino event is 6969.

In the case of the neutrino-EM coincidence events we found that for 12 (out of 24) events there are no galaxies within its 90 % containment error radius (*italic* in Table 2). The largest number of galaxies within 90 % containment error radius of a neutrino-EM coincidence event is 62. Within 50 % containment error radius also 12 events have no galaxies. The largest number of galaxies within 50 % containment error radius of a neutrino-EM coincidence event is 20.

**Table 1. The list of IceCube Gold alerts with the number of potential host galaxies of electromagnetic counterparts within 50 % and 90 % conatainment radii. For the neutrino events with galaxies within 50 % and 90 % conatainment radii, the name as Gal- (see Section 3) or ordering number of a closest galaxy from the Morphological catalog of SDSS galaxies [24] is presented in the last column**

| Event ID | RA, deg | Dec, deg | 50 % containment radius, arcmin | 90 % containment radius, arcmin | Number of galaxies within 50 % containment radius | Number of galaxies within 90 % containment radius | Closest galaxy within 90% containment radius |
|---|---|---|---|---|---|---|---|
| *141104_37015490* | 266.0000 | +38.9699 | 31.80 | 17.39 | 0 | 0 | - |
| *140870_26727884* | 59.6799 | +25.3200 | 28.05 | 14.39 | 0 | 0 | - |
| *140752_31006975* | 81.4699 | +16.9600 | 28.73 | 15.34 | 0 | 0 |  |
| *140626_1288692* | 211.0699 | -10.7300 | 17.54 | 8.69 | 0 | 0 | - |
| **140601_60511904** | **348.2200** | **+3.5499** | **27.00** | **15.00** | **1** | **2** | **Gal-nu-1** |
| 140125_41215060 | 164.0900 | +5.3799 | 31.80 | 17.39 | 18 | 47 | 102533* |
| *139977_2910365* | 112.3290 | -33.5285 | 52.23 | 18.59 | 0 | 0 | - |
| 139912_46959751 | 180.6599 | +18.9200 | 35.39 | 19.20 | 9 | 28 | 141702* |
| 139315_50057906 | 327.0799 | +3.0600 | 96.89 | 54.60 | 3 | 16 | 305597* |
| 139204_39158985 | 25.3999 | +7.7800 | 257.99 | 68.99 | 15 | 165 | 8752* |
| 139100_34742349 | 239.6299 | +39.9399 | 922.80 | 451.19 | 6969 | 27487 | 274357* |
| 138599_39138591 | 177.5300 | +53.6199 | 168.45 | 93.00 | 643 | 1940 | 134112* |
| *138515_8773328* | 105.6700 | +47.8500 | 147.59 | 86.99 | 0 | 0 | - |
| *138487_60138479* | 267.1600 | +46.9600 | 178.20 | 102.59 | 0 | 0 | - |
| 138415_56188508 | 143.7899 | +25.0399 | 65.40 | 35.99 | 17 | 98 | 63736* |
| *138283_14780365* | 17.9299 | -12.0999 | 17.99 | 10.20 | 0 | 0 | - |
| *138181_66037171* | 32.5200 | -1.8700 | 23.40 | 12.00 | 0 | 0 | - |
| *138131_35302784* | 270.6999 | +8.4600 | 201.00 | 115.19 | 0 | 0 | - |
| *138125_11333473* | 269.0299 | -1.9399 | 40.80 | 24.00 | 0 | 0 | - |
| *137910_29871391* | 50.1899 | +21.0599 | 187.20 | 108.60 | 0 | 0 | - |
| *137711_79205800* | 72.8599 | +34.2299 | 52.19 | 29.39 | 0 | 0 | - |
| 137487_35344578 | 31.8999 | +4.1799 | 82.20 | 46.20 | 9 | 23 | 11188* |
| *137467_64735045* | 350.5486 | +34.7110 | 39.58 | 15.41 | 0 | 0 | - |
| 136766_7637140 | 224.1200 | +41.3100 | 109.80 | 62.40 | 65 | 209 | 239679& |
| 136627_61640402 | 224.0300 | -1.3400 | 62.40 | 38.40 | 41 | 114 | 239507* |
| 136615_14688828 | 334.2500 | +5.3799 | 100.79 | 57.59 | 2 | 5 | 308067* |
| *136568_17854328* | 268.2400 | -10.7300 | 102.59 | 58.19 | 0 | 0 | - |
| *136565_2186969* | 346.1100 | +8.9100 | 68.39 | 37.80 | 0 | 0 | - |
| *136392_25495567* | 314.8199 | +8.6099 | 31.20 | 18.59 | 0 | 0 | - |
| 136388_4701751 | 48.7800 | +4.4800 | 372.00 | 226.79 | 53 | 827 | 14256* |
| *136385_7450363* | 267.8000 | +11.4199 | 70.79 | 41.99 | 0 | 0 | - |
| *136348_65788242* | 287.8399 | +20.7399 | 217.79 | 135.00 | 0 | 0 | - |
| *136260_4895987* | 266.8044 | -3.5750 | 30.80 | 12.00 | 0 | 0 | - |
| **136241_22093816** | **21.3599** | **-3.8799** | **43.79** | **27.00** | **1** | **3** | **Gal-nu-2** |
| **135908_43512334** | **225.9338** | **-0.2016** | **30.80** | **12.00** | **1** | **18** | **Gal-nu-3** |
| 135736_30987826 | 60.7299 | -4.1799 | 39.00 | 22.80 | 0 | 3 | 15713* |
| *135591_36044887* | 270.7900 | +25.2800 | 56.16 | 32.22 | 0 | 0 | - |
| *134994_1103075* | 155.2582 | -35.3994 | 71.40 | 22.94 | 0 | 0 | - |
| 134979_17138286 | 206.0600 | +4.7800 | 52.80 | 31.80 | 33 | 95 | 199550* |
| 134818_73718836 | 206.3700 | +13.4399 | 39.60 | 22.19 | 8 | 50 | 200417* |
| *134817_29175858* | 261.6899 | +41.8100 | 109.20 | 61.20 | 0 | 0 | - |
| 134777_8912764 | 6.8600 | -9.2500 | 65.40 | 38.40 | 31 | 79 | 2124* |
| *134751_31476488* | 30.5399 | -12.0999 | 70.79 | 41.39 | 0 | 0 | - |
| *134701_49508272* | 148.7519 | -21.6418 | 51.89 | 20.21 | 0 | 0 | - |
| 134699_70289682 | 195.1200 | +1.3799 | 77.40 | 42.59 | 70 | 180 | 17437*5 |
| *134698_40735501* | 105.2500 | +6.0499 | 64.80 | 34.19 | 0 | 0 | - |
| *134577_31638233* | 265.1700 | +5.3399 | 24.00 | 13.80 | 0 | 0 | - |
| 134552_68615710 | 29.5199 | +3.4700 | 31.80 | 24.00 | 2 | 3 | 10483& |
| *134533_53384881* | 96.4599 | -4.3300 | 39.60 | 23.40 | 0 | 0 | - |
| *134327_10548129* | 117.5554 | -24.8475 | 30.80 | 12.00 | 0 | 0 | - |
| 134191_17593623 | 142.9499 | +3.6600 | 73.19 | 43.19 | 80 | 191 | 62491** |
| 134139_35473338 | 255.3700 | +26.6099 | 160.19 | 65.40 | 53 | 539 | 297353* |
| 133609_37927131 | 164.4900 | +11.8699 | 174.60 | 96.60 | 392 | 1209 | 103423* |
| 133331_47828126 | 230.0999 | +3.1699 | 220.19 | 154.19 | 1651 | 3076 | 253593* |
| 133119_22683750 | 314.0799 | +12.9399 | 177.00 | 81.00 | 0 | 28 | 302403* |
| 133092_52499868 | 5.7599 | -1.5700 | 64.80 | 36.59 | 45 | 142 | 1799* |
| *133091_81419* | 167.4300 | -22.3900 | 177.00 | 101.39 | 0 | 0 | - |
| 132910_57145925 | 225.7899 | +10.4700 | 71.10 | 39.74 | 81 | 271 | 243328* |
| 132707_54984442 | 343.2599 | +10.7300 | 162.59 | 97.79 | 0 | 60 | 310612* |

\* Row number in the catalog [8]. where the SDSS Object ID is pointed out

**Table 2. The list of neutrino-EM coincidence events with the number of potential host galaxies of electromagnetic counterparts within 50% and 90% conatainment radii. For the coincidence events with galaxies within 50 % and 90 % conatainment radii, the name Gal- (see Section 3) or ordering number of a closest galaxy from the Morphological catalog of galaxies [24] is presented in the last column**

| Event ID | RA, deg | Dec, deg | 50% containment radius, arcmin | 90% containment radius, arcmin | Number of galaxies within 50% containment radius | Number of galaxies within 90% containment radius | Closest galaxy within 90% containment radius |
|---|---|---|---|---|---|---|---|
| *0_189499* | 93.8299 | +21.0199 | *20.29* | *11.40* | 0 | 0 | - |
| **0_186631** | **164.7100** | **+23.5199** | **14.04** | **7.80** | **1** | **6** | **Gal-nuEM-1** |
| 0_180410 | 253.5600 | +29.6400 | 18.89 | 10.20 | 5 | 22 | 295747* |

| | | | | | | | | |
|---|---|---|---|---|---|---|---|---|
| 0_179815 | 212.8100 | +17.2100 | 26.95 | 15.00 | 4 | 15 | 215156* | |
| 0_164965 | 149.9699 | +46.2199 | 24.34 | 13.20 | 3 | 15 | 74633& | |
| *0_149605* | *251.7299* | *+53.7800* | *32.86* | *17.99* | *0* | *0* | *-* | |
| *0_142904* | *313.1800* | *+7.2000* | *18.59* | *10.20* | *0* | *0* | *-* | |
| **0_141028** | **331.9200** | **+12.4399** | **14.11** | **7.80** | **1** | **2** | **Gal-nuEM-2** | |
| **0_134980** | **143.4000** | **+4.8300** | **15.29** | **8.40** | **1** | **5** | **Gal-nuEM-3** | |
| *0_132619* | *47.2199* | *+27.4200* | *28.02* | *15.60* | *0* | *0* | *-* | |
| 0_130124 | 206.0099 | +16.1600 | 14.05 | 7.80 | 3 | 10 | 199590* | |
| 0_127938 | 198.2299 | +59.5099 | 25.18 | 13.80 | 5 | 20 | 182021* | |
| *0_121306* | *108.8799* | *+40.9200* | *29.77* | *16.20* | *0* | *0* | *-* | |
| 0_108943 | 221.3499 | +13.2300 | 18.92 | 10.20 | 6 | 14 | 233638* | |
| *0_108094* | *307.5699* | *+1.6000* | *34.31* | *18.59* | *0* | *0* | *-* | |
| *0_105322* | *322.1299* | *+27.2600* | *15.68* | *8.40* | *0* | *0* | *-* | |
| *0_101674* | *12.0299* | *-5.7500* | *19.26* | *10.79* | *0* | *0* | *-* | |
| *0_96720* | *99.7600* | *+9.0700* | *18.74* | *10.20* | *0* | *0* | *-* | |
| *0_85791* | *93.9300* | *+12.5099* | *21.44* | *12.00* | *0* | *0* | *-* | |
| *0_85790* | *93.6400* | *+14.6600* | *16.06* | *8.99* | *0* | *0* | *-* | |
| 0_73310 | 162.3400 | +19.4600 | 40.19 | 22.19 | 20 | 62 | 98995* | |
| 0_68186 | 134.9900 | +7.7400 | 25.10 | 13.80 | 11 | 25 | 49014* | |
| 0_66291 | 140.1999 | +29.7600 | 16.30 | 8.99 | 2 | 7 | 58139* | |
| *0_54519* | *118.5000* | *-1.6200* | *22.76* | *12.59* | *0* | *0* | *-* | |

*row number in the Catalog [8], where the SDSS Object ID is pointed out

## 4. Discussion and conclusion

The most prospective are 3 neutrino events (**bold** in Table 1) and 3 neutrino-EM coincidence events (**bold** in Table 2) with only 1 galaxy within their 50 % containment radii. We cross-matched various multiwavelength sky surveys to verify their status as the potential host galaxy candidates to be source of the neutrino event ID / neutrino-EM coincidence event ID and collected the data in Tables 3 and 4 (we used the HyperLEDA [12] and NED databases for the multiwavelemgth properties).

**Table 3. Data on the neutrino event ID / neutrino-EM coincidence event ID and potential host SDSS galaxy**

| Event ID name in Tables 1 and 2 | Event discovery date | SDSS galaxy name | SDSS galaxy RA, Dec, in deg | SDSS galaxy redshift, cosmological distance modulus*, total B-mag | SDSS galaxy type | SDSS galaxy feature | SDSS galaxy in ZTF, transient light curve |
|---|---|---|---|---|---|---|---|
| **The neutrino event ID and potential host SDSS galaxy** | | | | | | | |
| 140601_60511904 Gal-nu-1 | 2025 March 2 | **J231231.52+033415.1** | 348.1313 +3.570 | 0.068 37.42 18.37 | Spirsl | Round – in between | + |
| 136241_22093816 Gal-nu-2 | 2022 Feb 2 | **J012546.81-032938.7** | 21.4450 -3.4940 | 0.087 37.97 17.94 | Elliptical (?) | - | + |
| 135908_43512334 Gal-nu-3 | 2021 Nov 17 | **J150337.60-002332.2** | 225.9066 -0.3922 | 0.037 36.06 17.80 | Spiral | Disk | + |
| | | | | | | | |

| The neutrino-EM coincidence event ID and potential host SDSS galaxy | | | | | | | |
|---|---|---|---|---|---|---|---|
| 0_186631 Gal-nuEM-1 | 2025, July 30 | J105846.51+232742.2 | 164.6938 23.4617 | 0.048 36.64 16.72 | Spiral Sbc | Two spiral arms, ring | + |
| **0_141028 Gal-nuEM-2** | **2023, Sep 27** | **J220711.14+122535.9** | **331.7964 12.4266** | **0.079 37.76 18.37** | **Spiral** | **AGN** | **+** |
| 0_134980 Gal-nuEM-3 | 2023, May 23 | J093317.17+045157.5 | 143.3215 4.8659 | 0.076 16.666 | Spiral | Disk | + |

*The redshift distance modulus (in *mag*) is defined as $modz = 5 \log(D_L) + 25$, where $D_L$ is the luminosity-distance (in Mpc) defined by the law of dimming of the observed flux: $F \propto L / D_L^2$., ΛCDM parameters $H_0 = 70$ km/s/Mpc, $\Omega_M = 0.27$, $\Omega_\Lambda = 0.73$.

**Table 4. Multi-wavelength data on potential host galaxies of a source of the neutrino event ID / neutrino-EM coincidence event ID**

| | |
|---|---|
| 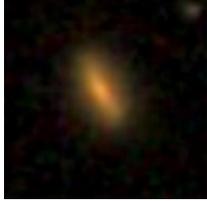 | **SDSS J231231.53+033415.1** VisS. Cross-matching with other catalogs: WISEA J231231.51+033415.1 IrS, 2MASX J23123151+0334152 ID G, 2MASX J23123151+0334152 IrS, 2MASS J23123151+0334152 IrS It's observed only in optical and infrared sky surveys. There are no other references to this galaxy. **The light curve taken with the ZTF data** is in Fig, 1, |
| 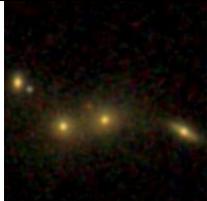 | **SDSS J012546.81-032938.7** VisS. Cross-matching with other catalogs: WISEA J012546.79-032938.8 IrS, 2MASXi J0125468-032938 IrS, 2MASX J01254681-0329381 IrS, 2MASS J01254679-0329384 IrS It's observed only in optical and infrared sky surveys. There are no other references to this galaxy. |
| 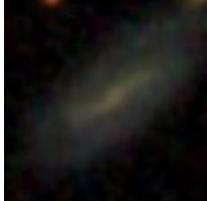 | **SDSS J150337.59-002332.2** (MaNGA 01-014340 G,(PGC1144481), Cross-matching with other catalogs: WISEA J150337.57-002331.8 IrS, GALEXMSC J150337.59-002330.6 UvS, , NSA 002713 G, LEDA 1144481 G, GALEXASC J150337.58-002330.3 UvS, [TTL2012] 143240 G. It's observed in ultraviolet, optical, infrared ranges; it was included into the Automatic Spectroscopic K-means-based classification (ASK 008680.0). This galaxy is a member of merging group [21, 28] with many star-forming regions [7], when group is a part of filamentary structure of cosmic web [20] |
| 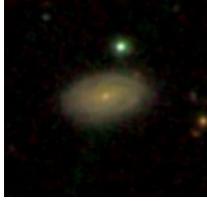 | **SDSS J105846.51+232742.2** (MaNGA 01-486291 G, SDSS J105846.51+232742.3 G). Cross-matching with other catalogs: WISEA J105846.51+232742.4 IrS, 2MASXi J1058464+232742 IrS, MAPS-NGP O_374_0405546 G, 2MASS J10584649+2327421 IrS, NSA 111460 G, GALEXASC J105846.48+232743.5 UvS, [TTL2012] 103047 G. It's observed in ultraviolet, optical, near and mid infrared, ranges; it was included into the Automatic Spectroscopic K-means-based classification (ASK 619867.0). This is a typical spiral galaxy [26], a member of merging group [21, 28] with many star-forming regions [7], when group is a part of filamentary structure of cosmic web [20].. |
| 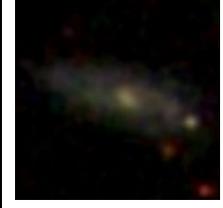 | **SDSS J220711.14+122535.8** G (SDSS J220711.14+122535.9 G). Cross-matching with other catalogs: WISEA J220711.10+122535.6 IrS, GALEXASC J220711.09+122536.9 UvS, GALEXMSC J220711.14+122535.9 UvS, INTEGRAL It's observed in ultraviolet, optical, near and mid infrared, ranges, has many star-forming regions [7]. It was included into the Automatic Spectroscopic K-means-based classification (ASK 140252.0). **The light curve taken with the ZTF data** is in Fig, 1, |
| 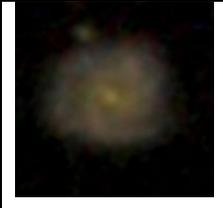 | **SDSS J093317.15+045157.5** G (SDSSCGB 39458.01 G). Cross-matching with other catalogs: WISEA J093317.15+045157.5 IrS, GALEXMSC J093317.16+045159.5 UvS, GALEXASC J093317.14+045159.8 UvS. It's observed in ultraviolet, optical, near and mid infrared ranges, a member of galaxy group [21], has many star-forming regions [7]. It was included into the Automatic Spectroscopic K-means-based classification (ASK 228158.0). |

All these galaxies were observed in ultraviolet (GALEX), optical (mostly SDSS), and infrared (WISE, 2MASS) sky surveys. They are the faint objects having no a strictly defined morphological type, spectral activity type of its nuclei, and multiwavelength features.

In order to search for optical counterparts to six IceCube neutrino events in the ZTF alert stream, we queried the ZTF Archive using AMPEL's streaming interface. For each event, we defined a circular search region centered on the equatorial coordinates (RA, Dec) with a radius set to the IceCube 90% containment. The temporal window spanned from T0−7 days to T0+14 days around the IceCube trigger time T0 (expressed as Modified Julian Date). For each event, we submitted a cone search to the ZTF Archive API (v3) via the streams from_query end point, and iterated over all matching alerts using ZTF Archive Alert Loader. As a result, across all six searches, the loader returned zero alerts within the specified cones and time windows for these neutrino events.

The light curves of these galaxies taken from ZTF have some photometric peculiarities, but don't reveal the transient features, which can be also explained that they are less than signal/noise ratio.

The light curve of potential host galaxies SDSS J231231.52+033415.1 and SDSS J220711.14+122535.9 as the source of neutrino event / neutrino-EM coincidence event demonstrates are presented in Fig. 1. They are spiral galaxies with active nuclei, similar total B-magnitude as $18^m.37$; they both are distant galaxies in the Local Volume with redshifts as $z = 0.068$ and $z = 0.079$, cosmological distance moduli as 37.42 mag and 37.76 mag ($H_0 = 70$ km/s/Mpc), respectively.

The open ZTF database was used to construct light curves in the *g,r,i* filters (*ugriz system*) and color indices *g-r* (relative to the *g* filter) and *r-i* (relative to the *r* filter) for the period from 2018 to 2025. The York method [29] was used for linear regression, where the slope of the regression line (*m*) in the York method describes how the color index changes with respect to brightness of a galaxy. It takes into account measurement errors on both the X and Y axes, and correlation between them. This allows for more accurate estimates of regression parameters, including the slope of the regression line (*m*). Additionally, Pearson correlation coefficients (*r*) were computed to analyze the relationship between the quantities. For the object SDSS J220711.14+122535.9 (Fig. 1, right panel), a medium positive correlation $r(r–i) = 0.642$ and a high positive correlation $r(g–r) = 0.746$ were found, and for the object SDSS J231231.52+033415.1 (Fig. 1, left panel), a high positive correlation $r(g-r) = 0.763$ was found. Fig.1 shows the dependence between the brightness in the *g* filter and the color index *g-r*, as well as between the *r* filter and the color index *r-i*; the DR(*N*) parameter is a number of points on the graph, and *p*-value gives a statistical significance of the regression line's slope. All color indices show a Bluer-when-brighter trend (BWB) for these galaxies..

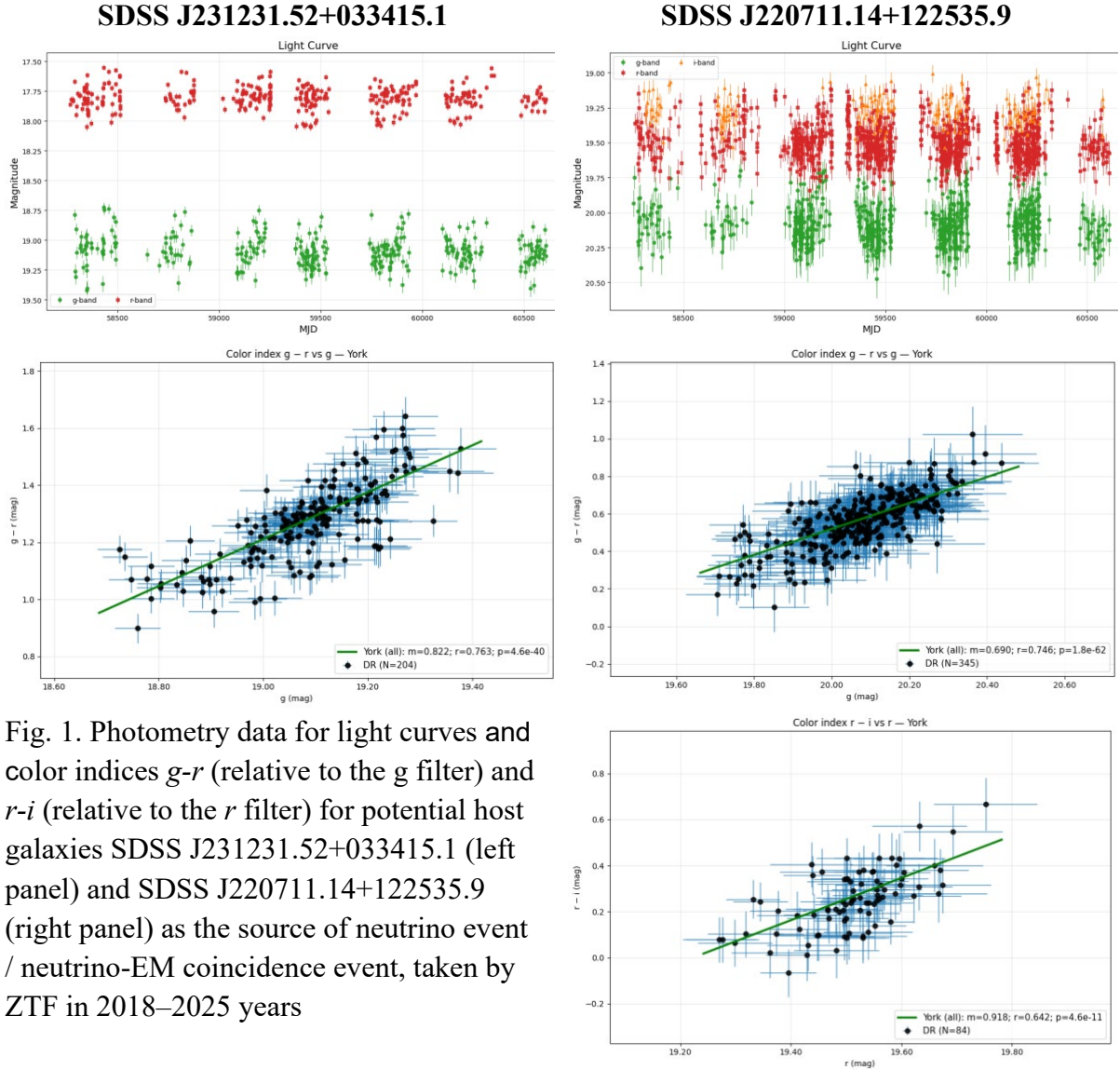

Fig. 1. Photometry data for light curves and color indices *g-r* (relative to the g filter) and *r-i* (relative to the *r* filter) for potential host galaxies SDSS J231231.52+033415.1 (left panel) and SDSS J220711.14+122535.9 (right panel) as the source of neutrino event / neutrino-EM coincidence event, taken by ZTF in 2018–2025 years

So, we searched for spatial coincidence between the galaxies from SDSS and the high-energy neutrino events. The IceCube Gold alerts and the neutrino-electromagnetic coincidence events from AMON (Astrophysical Multimessenger Observatory Network) identified until the end of September 2025 are considered. Galaxies from the Morphological catalog of galaxies at 0.02<z<0.1 (including 315,776 objects from SDSS DR9 with the absolute stellar magnitudes in the range of $-24^m < M^r < -13^m$) are examined. Among 59 IceCube Gold alerts we ound 3 with only 1 galaxy (SDSS J231231.52+033415.1) within the 50 % containment radius. Among 24 neutrino-electromagnetic coincidence events there are also 3 with only 1 galaxy (SDSS

J220711.14+122535.9) within the 50 % containment radius. These 6 galaxies are the most promising candidates for host galaxies of the neutrino sources. The available multiwavelength data in ultraviolet (GALEX), optical (mostly SDSS), and infrared (WISE, 2MASS) sky surveys shows that these galaxies are very poorly studied. They are the faint objects having no a strictly defined morphological type, spectral activity type of its nuclei. We submitted a cone search to the ZTF Archive API (v3) via the streams from_query end point, and iterated over all matching alerts using ZTF Archive Alert Loader. As a result, the loader returned zero alerts within the specified cones and time windows for these neutrino events. The light curves of these galaxies taken from ZTF have some photometric peculiarities, namely, all color indices show a Bluer-when-brighter trend (BWB), but don't reveal any transient features, which can be also explained that they are less than signal/noise ratio.

**Acknowledgements**. This work is conducted in frame of the project 2023.03/0188 of the National research Fund of Ukraine. The research by Izviekova I.O. is a part of the project (ID 848), which is conducted in the frame of the EURIZON program funded by the European Union under grant agreement No.871072. We thank Prof. M. Kowalski, Dr. J. Nordin and Dr. J. van Santen (DESY, Humboldt University, Germany) for fruitful seminar and discussions in March 2025, when the results of this research were presented as well as Dr. Dobrycheva D.V. for her helpful remarks. We acknowledge the usage of the AMON and ZTF web-sites as well as the NASA/IPAC Extragalactic Database, which is funded by the NASA and operated by the California Institute of Technology, and HyperLEDA database (http://leda.univ-lyon1.fr).. ..

**References**

1. Aartsen M. G., Ackermann M., Adams J. et al (2017), The IceCube real-time alert system. Astroparticle Physics, **92**, pp. 30–41. https://doi.org/10.1016/j.astropartphys.2017.05.002

2. Aartsen M., Ackermann M., Adams J. (2018). Neutrino emission from the direction of the blazar TXS 0506+056 prior to the IceCube-170922A alert. Science, **361**, pp. 147–151. https://doi.org/10.1126/science.aat2890

3. Abbasi R., Ackermann M., Adams J. et al (2022). Evidence for neutrino emission from the nearby active galaxy NGC 1068. Science, 378, 538–543. https://doi.org/10.1126/science.abg3395

4. Ayala Solares H;A., Coutu S., Cowen D.F. et al. (2020_. The Astrophysical Multimessenger Observatory Network (AMON): Performance and science program. Astroparticle Physics, **114**, pp. 68–76. https://doi.org/10.1016/j.astropartphys.2019.06.007

5. Blaufuss E., Kintscher T., Lu L., and Tung C. F. The Next Generation of IceCube Real-time Neutrino Alerts (2020). In 36th International Cosmic Ray Conference (ICRC2019). Proceedings of Science, ICRC2019, **36**, Art. no 1021. https://doi.org/10.22323/1.358.01021

6. Burgazli, A., Sergijenko, O., Vavilova, I. (2022). Machine learning in cosmology and gravitational wave astronomy: recent trends. In book "Horizons in Computer Science Research, **22**, pp. 193–240. New York, Nova Science Publisher Inc. ISBN 979-8-88697-101-9, 979-8-88697-234-4(e-Book)

7. Chang Y.-Y., van der Wel A., da Cunha E., and Rix H.-W. (2015). Stellar Masses and Star Formation Rates for 1M Galaxies from SDSS+WISE. Astrophys. J. Suppl. Ser., **219**, no. 1, Art. no. 8. https://doi,oeg/10.1088/0067-0049/219/1/8

8. Duev D.A., Mahabal A.; Masci F. J. et al. (2019). Real-bogus classification for the Zwicky Transient Facility using deep learning. Mon. Notic. R. Astron. Soc., **489**, no. 3. pp. 3582–3590, https://doi.org/:10.1093/mnras/stz2357

9.. IceCube Collaboration (2013). Evidence for High-Energy Extraterrestrial Neutrinos at the IceCube Detector. Science, **342**, 1242856 https://doi.org/10.1126/science.124285

10. IceCube Collaboration (2023). Observation of high-energy neutrinos from the Galactic plane. Science, **380**, 1338-1343. https://doi.org/10.1126/science.adc9818

11. Khramtsov, V., Vavilova, I.B., Dobrycheva, D.V. et al. (2022). Machine learning technique for morphological classification of galaxies from the SDSS. III. The CNN image-based inference of detailed features. Space Sci. & Technol.; 28(5), p. 27–55 https://doi.org/10.15407/knit2022.05.027

12. Makarov, D., Prugniel, P., Terekhova, N., Courtois, H., and Vauglin, I. (2014). HyperLEDA. III. The catalogue of extragalactic distances. Astron. Astrophys., vol. 570, Art. no. A13. https://doi.org/10.1051/0004-6361/201423496

13. Neronov A., Semikoz D., Savchenko D.(2024). Neutrino signal from Cygnus region of the Milky Way. Phys. Rev. D, **110**, 043024. https://doi.org/10.1103/PhysRevD.110.043024

14. Nordin, J., Brinnel, V., van Santen, J. et al. (2019). Transient processing and analysis using AMPEL: Alert management, photometry, and evaluation of light curves. Astron. Astrophys., **631**, A147. https://doi.org/10.1051/0004-6361/201935634

15. Nordin, J., Brinnel, V., van Santen, J., Reusch, S., and Kowalski, M. (2025). AMPEL workflows for LSST: Modular and reproducible real-time photometric classification, Astron. Astrophys., **698**, Art. no. A13, EDP. https://doi,org/10.1051/0004-6361/202452481

16. Reusch S., Stein R., Kowalski M. et al. (2022). Candidate Tidal Disruption Event AT2019fdr Coincident with a High-Energy Neutrino. Phys. Rev. Lett.. **128**, 221101,: https://doi.org/10.1103/PhysRevLett.128.221101.

17. Sergijenko O., Brown A., Fiorillo D. et al. (2021). Sensitivity of the Cherenkov Telescope Array to emission from the gamma-ray counterparts of neutrino events. Proceedings of Science, ICRC2021, Art. no 975. PoS(ICRC2021)975. https://doi.org/10.22323/1.395.0975

18. Sergijenko O., Brown A., Fiorillo D. et al. (2023). Sensitivity of the Cherenkov Telescope Array to the gamma-ray emission from neutrino sources detected by IceCube Proceedings of Science, ICRC2023, Art. no 1531. PoS(ICRC2023)1531 https://doi.org/10.22323/1.444.1531

19. Stein, R., Velzen, S.v., Kowalski, M. et al. (2021). A tidal disruption event coincident with a high-energy neutrino. Nat. Astron., **5**, pp. 510–518 https://doi.org/10.1038/s41550-020-01295-8

20. Tempel, E., Stoica, R. S., Martínez, V. J. et al. (2014). Detecting filamentary pattern in the cosmic web: a catalogue of filaments for the SDSS. Mon. Notic. R. Astron. Soc., **438**, no. 4, pp. 3465–3482. https://doi:org/10.1093/mnras/stt2454

21. Tempel, E., Tuvikene, T., Kipper, R., and Libeskind, N. I. (2017). Merging groups and clusters of galaxies from the SDSS data. The catalogue of groups and potentially merging

systems.Astron. Astrophys., **602**, Art. no. A100. https://doi.org/10.1051/0004-6361/201730499

22. Vavilova I. B., Dobrycheva D.V., Vasylenko M. Yu. (2021), Machine learning technique for morphological classification of galaxies from the SDSS. I. Photometry-based approach, Astron. Astrophys., **648**, p. A122, https://doi.org/10.1051/0004-6361/202038981

23. Vavilova I.B., Khramtsov V., Dobrycheva D.V. et al. (2022). Machine learning technique for morphological classification of galaxies from SDSS. II. The image-based morphological catalogs of galaxies at 0.02<z<0.1 . Space Sci. & Technol., **28**(1), p. 3-22. https://doi.org/10.15407/knit2022.01.003

24. Vavilova I.B., Khramtsov V. , Dobrycheva D.V. (2023). VizieR Online Data Catalog: Galaxies at 0.02<z<0.1 morphological catalog. https://cdsarc.cds.unistra.fr/viz-bin/cat/J/other/KNIT/28.3

25. Vavilova, I. B. ; Dobrycheva, D. V. ; Khramtsov, V. et al. (2024). Machine Learning of Galaxy Classification by their Images and Photometry. Astronomical Society of the Pacific Conference Series, 535 , p.103–107.

26. Wei, J. (2024).A New Statistical Analysis of the Morphology of Spiral Galaxies. Astron. J., **168**, no. 6, Art. no. 264. https://doi.org/10.3847/1538-3881/ad8632

27. Willett K. W., Lintott C. J., Bamford S. P., et al. (2013). Galaxy Zoo 2: detailed morphological classifications for 304 122 galaxies from the Sloan Digital Sky Survey. Mon. Notic. Roy. Astron. Soc., **435**, no.4, p. 2835–2860. https://doi.org/10.1093/mnras/stt1458.

28. Yang, X., Mo, H. J., van den Bosch, F. C., Pasquali, A., Li, C., and Barden, M. (2007). alaxy Groups in the SDSS DR4. I. The Catalog and Basic Properties. Astrophys. J., vol. 671, no. 1, pp. 153–170. https://doi.org/10.1086/522027.

29. York D., Evensen N.M., Martínez M.L., De Basabe D.J. (2004). Unified equations for the slope, intercept, and standard errors of the best straight line. American Journal of Physics, **72**, 3, pp.367–375. https://doi.org/10.1119/1.1632486